\documentclass[%
twocolumn,
groupedaddress,
nofootinbib,
bibnotes,
 amsmath,amssymb,aps
]{revtex4-2}
\usepackage{xcolor}
\usepackage{graphicx}
\usepackage{dcolumn}
\usepackage{bm}

\begin{document}
\preprint{APS/123-QED}

\title{An effective theory of collective deep learning} 

\author{Llu\'is Arola-Fern\'andez}
\email{lluisarolaf@gmail.com}

\affiliation{Instituto de F\'isica Interdisciplinar y Sistemas Complejos IFISC (CSIC-UIB),\\Campus UIB, 07122 Palma de Mallorca, Spain}%

\author{Lucas Lacasa}
\email{lucas@ifisc.uib-csic.es}

\affiliation{Instituto de F\'isica Interdisciplinar y Sistemas Complejos IFISC (CSIC-UIB),\\Campus UIB, 07122 Palma de Mallorca, Spain}%

\date{\today}

\begin{abstract}

Unraveling the emergence of collective learning in systems of coupled artificial neural networks points to broader implications for machine learning, neuroscience, and society. Here we introduce a minimal model that condenses several recent decentralized algorithms by considering a competition between two terms: the local learning dynamics in the parameters of each neural network unit, and a diffusive coupling among units that tends to homogenize the parameters of the ensemble. We derive an effective theory for linear networks to show that the coarse-grained behavior of our system is equivalent to a deformed Ginzburg-Landau model with quenched disorder. This framework predicts depth-dependent disorder-order-disorder phase transitions in the parameters’ solutions that reveal a depth-delayed onset of a collective learning phase and a low-rank microscopic learning path. We validate the theory in coupled ensembles of realistic neural networks trained on the MNIST dataset under privacy constraints. Interestingly, experiments confirm that individual networks --trained on private data-- can fully generalize to unseen data classes when the collective learning phase emerges. Our work establishes the physics of collective learning and contributes to the mechanistic interpretability of deep learning in decentralized settings. 

\end{abstract}

\maketitle

Collective behavior emerging from the dynamics of many interacting particles or units is the flagship of complexity \cite{anderson72,mezard86,strogatz03} and a common feature found across natural and artificial complex systems \cite{munoz18,sole19,pinero19}. 
Learning, a paradigmatic example of collective behavior \cite{hebbs49,mataric93}, is a fascinating ability of the brain, where large ensembles of neurons interact, adapting their synaptic circuits in such a way that allow us to learn from experience. Learning can also occur in a collective manner among interacting agents that learn from each other, and at different scales, from ant colonies \cite{sole19} to robot swarms \cite{ha22} or social communities \cite{olsson20}. Such generality and multiscale nature of learning has attracted a great interest across fields and the major advances in the problem have definitely benefited from a cross-disciplinary research. In fact, ideas from complex systems and statistical physics are at the roots of the early models of computation in neural networks \cite{hopfield82} and in classical machine learning tools such as random forests and particle swarm optimization \cite{kennedy95}, among other notable, more recent contributions \cite{carleo2019machine,mezard23,liu22grokking,ziyin2022exact,mehta19,roberts22}. 

The many-particle approach has proved particularly successful in deep learning \cite{goodfellow16}, a collection of algorithms and techniques involving very large and overparametrized neural networks, which has recently shown astonishing results in a myriad of challenging tasks \cite{jumper21, brown19} and intriguing emerging behavior in large language models \cite{bubeck2023sparks}. Interestingly, industry-related constraints posed by e.g. data privacy issues and power consumption during training \cite{bennun20,li19}, along with the perpetual quest for finding architectures with enhanced performance, have driven the field of deep learning to explore new solutions that, in hindsight, capitalize on some form of collective behavior \cite{goodfellow16,mehta19}. These range from ensemble approaches \cite{lee15,bennun20} to other decentralized solutions including federated learning \cite{mcmahan17,li19,liang20}, cooperative learning \cite{wang21} or transfer learning \cite{zhuang19}. This engineering success provides a strong motivation to understand, from a principled complex-systems viewpoint, whether collections of interacting `brains' --rather than interacting `neurons'-- do indeed develop collective behavior in the physical sense \cite{anderson72,sole19,mehta19}. Does collective learning emerge when neural networks are put in interaction? Does this collective learning phase emerge abruptly --as in the theory of phase transitions--? And do deep learning architectures play a non-trivial role in such phenomenology? In this Letter we provide affirmative answers to these questions. We present a minimal mathematical model of collective learning where we show that local brains solely trained for isolated tasks (private data) can generalize far outside their training set when coupled, and this happens via the onset of a collective learning phase transition. Our results are predicted by a physical effective theory (amenable to mechanistic interpretability \cite{carleo2019machine,mezard23,liu22grokking,ziyin2022exact}) and subsequently confirmed in a range of realistic experiments.

\begin{figure}[]
  \includegraphics[width=\columnwidth]{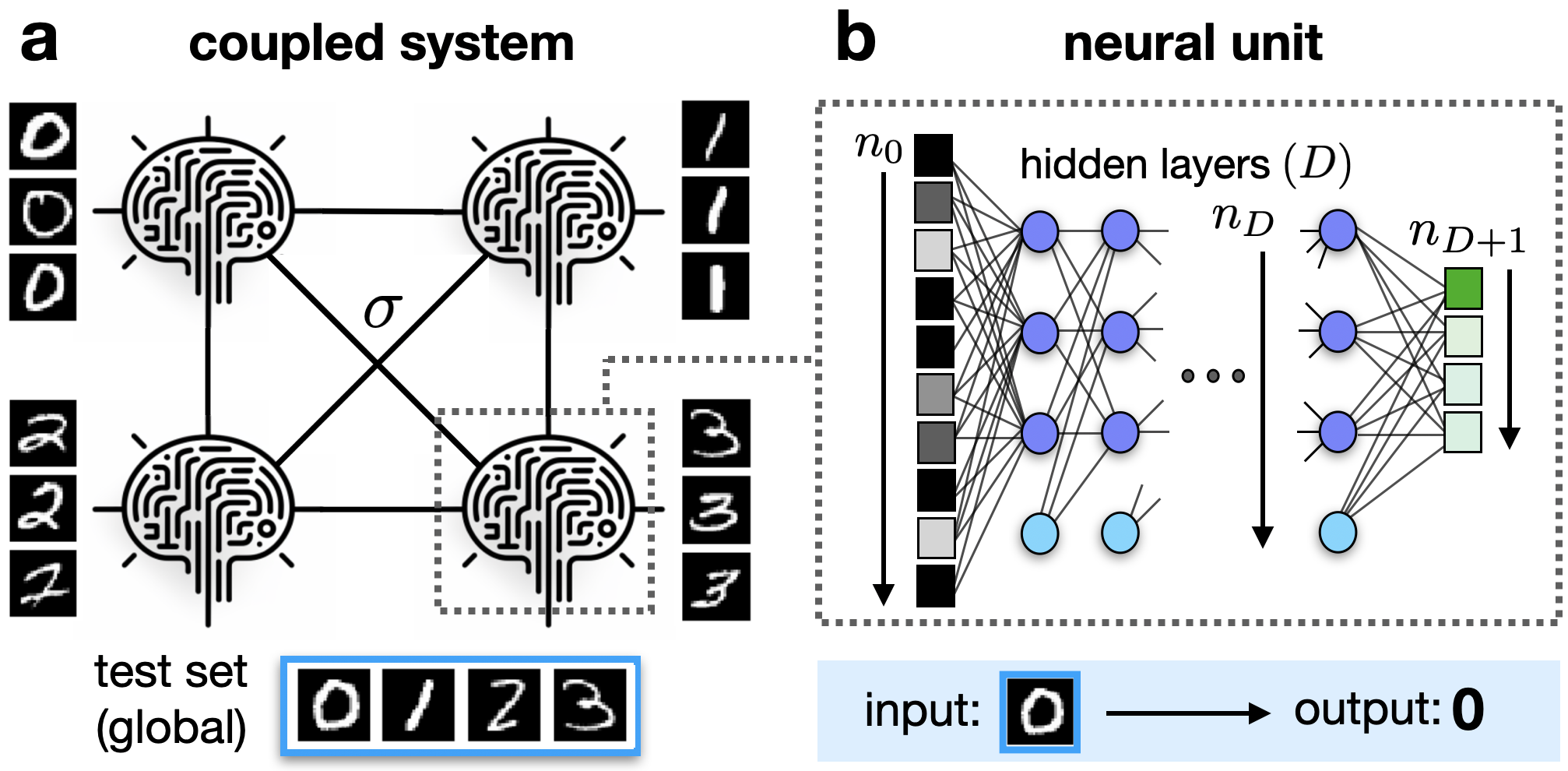}
  \caption{\textbf{Collective learning in coupled neural networks: } \textbf{a}: Small mean-field network of $N = 4$ coupled neural units trained on private data classes of MNIST, interacting with coupling strength $\sigma$. \textbf{b}: Simplified deep feed-forward architecture of a neural unit (trained on digit-3 images). A test point far outside the training set (a digit-0 image) is flattened as an input and processed towards the output layer. Here the unit predicts well an unseen class during training due to a collective learning mechanism emerging from the interactions.}
  \label{fig1}
\end{figure}

\noindent {\bf Collective learning model --} Our proof of concept considers a classification task to be solved by a coupled ensemble of \emph{vanilla} feed-forward neural networks, where each neural network is trained on data from a single class and evaluation is performed on the whole (multiclass) test set. However, the framework introduced below is 
flexible and extends to a generic supervised task or neural architecture. Let $\cal{D}=\{(\bf{x, y})\} \subseteq \ $${\mathbb{R}}$$^{{n_0}} \times {\mathbb{R}}^{n_{D+1}}$ denote the training set, where $\cal{X} = \{{\bf x}\ : ({\bf x,y}) \in \cal{D}\}$ and $\cal{Y} = \{{\bf y}\ : ({\bf x,y}) \in \cal{D}\}$ denote the input and output (labels) vectors with dimension $n_0$ and $n_{D+1}$, respectively. The training data is partitioned and distributed across $N$ neural units, such that ${\cal D}=\cup_{i=1}^N {\cal D}_i$. Each unit is a fully-connected feed-forward deep network with $D$ hidden layers with widths $n_d$, for $d = 1,\dots,D$, a readout (output) layer with $n_{D+1}$ neurons, and nonlinear activation functions in the hidden neurons. We define $\bar{\bf{y}} = f_i({\bf x},\boldsymbol{\theta}_i)$ as the predicted output value given the input $\bf x$ and the set of trainable parameters $\boldsymbol{\theta}_i=\{\theta_i^\alpha\}$, where $\alpha$ runs over all parameters of the $i$-th unit. We focus on a learning process under privacy constraints, such that each neural unit seeks to minimize the local empirical loss function
\begin{equation}
    {\cal{L}}_i = \sum_{({\bf x}_i,{\bf y}_i) \in {{\cal{D}}}_i} \ell[f({\bf x}_i,\boldsymbol{\theta}_i),{\bf y}_i]+ \gamma||\boldsymbol{\theta}_i||_2, 
    \label{eq:NNloss}
\end{equation}
where the sum runs over the data tuples associated to the $i$-th unit. The individual loss function $\ell(\bar{\bf{y}}_i,\mathbf{y}_i) : \mathbb{R}^{n_{D+1}} \times \mathbb{R}^{n_{D+1}} \rightarrow \mathbb{R}$, measures the error between the prediction on a data point and the corresponding true value (or label) and the right-hand term is a standard weight decay ($L_2$ regularization) with strength $\gamma$ that encourages a lower model complexity, improving generalization \cite{goodfellow16,mehta19}. To minimize Eq.(\ref{eq:NNloss}), the Backpropagation algorithm tunes the parameters of the units efficiently via Stochastic Gradient Descent (SGD) \cite{goodfellow16}. Since local learning is restricted to the private data of each unit, we introduce an interaction mechanism to induce collective learning in the system, as explained in Fig. \ref{fig1}. As a minimal mechanism, we select a consensus-based model \cite{wang21}, where units are diffusively coupled in a relation `parameter-to-parameter'. The interaction links (and their intensities) are captured by the weighted adjacency matrix $Q \in \mathbb{R}^{N \times N}=\{q_{ij}\}$ of the supra-network. Combining both local and interaction terms, each of the $\alpha$ parameters of a unit is updated at each iteration with
\begin{equation}
    \theta^{\alpha}_{i}(t+1) = \theta^{\alpha}_{i}(t)  -\eta\nabla_{\theta^{\alpha}_{i}}  {\cal{{L}}}_i
     + \frac{\eta \sigma}{N} \sum_{j = 1}^N q_{ij} \left(\theta^{\alpha}_{j}(t)-\theta^{\alpha}_{i}(t)\right), 
    \label{eq:NNdynamics}
\end{equation}
where $\eta$ is the learning rate and $\sigma$ the coupling strength of the interactions. To assess collective learning, we let the dynamics relax to a stationary state and then measure the mean test loss, i.e. the local loss evaluated on global (multiclass) test data $\cal{D}^{\text{test}}$, averaged over the units
\begin{equation}
    \langle L \rangle = \frac{1}{N} \sum_{i=1}^N \sum_{({\bf x,y}) \in {\cal{D}}^{\text{test}}} \ell[f({\bf x},\boldsymbol{\theta}_i),{\bf y}],
    \label{eq:NNloss1}
\end{equation}
which serves as an order parameter that measures the collective performance of the system (we expect a low loss as a fingerprint of collective learning, indicating that units can generalize outside their training set). In classification tasks with discrete classes, the accuracy metric $\langle A \rangle$ measures the percentage of correct class predictions in the test set (again brackets indicate an average over neural units). To capture the microscopic details of the process, we use the cross-loss and cross-accuracy matrices $L_x, \ A_x \in \mathbb{R}^{N \times N}$, where the entry $(i,j)$ determines the loss (accuracy) of the $i$-th unit when evaluated on data classes assigned to the $j$-th unit.
Note that the dynamics in Eq.(\ref{eq:NNdynamics}) lies in the cooperative learning framework \cite{wang21}, akin to decentralized schemes such as elastic averaging SGD \cite{zhang2015deep} and also captures the spirit of federated learning \cite{mcmahan17}, with the conceptual difference that here the self-organized dynamics of the units is independent of a centralized \emph{master} model. Previous results proved the convergence of cooperative and federated schemes \cite{zhang2015deep,liang20,woodworth20,wang21,li22,baihe21,li2019convergence}, unveiling that while heterogeneous and private data tend to slow down convergence \cite{li2019convergence}, distributed schemes can find solutions with higher generalization properties \cite{zhang2015deep,baldassi16,baldassi20}. The mechanisms underpinning the onset of emergent collective behavior in coupled neural networks under privacy constraints remain, however, poorly understood.

\medskip 

\begin{figure}[h]
  \centering
  \includegraphics[width=\columnwidth]{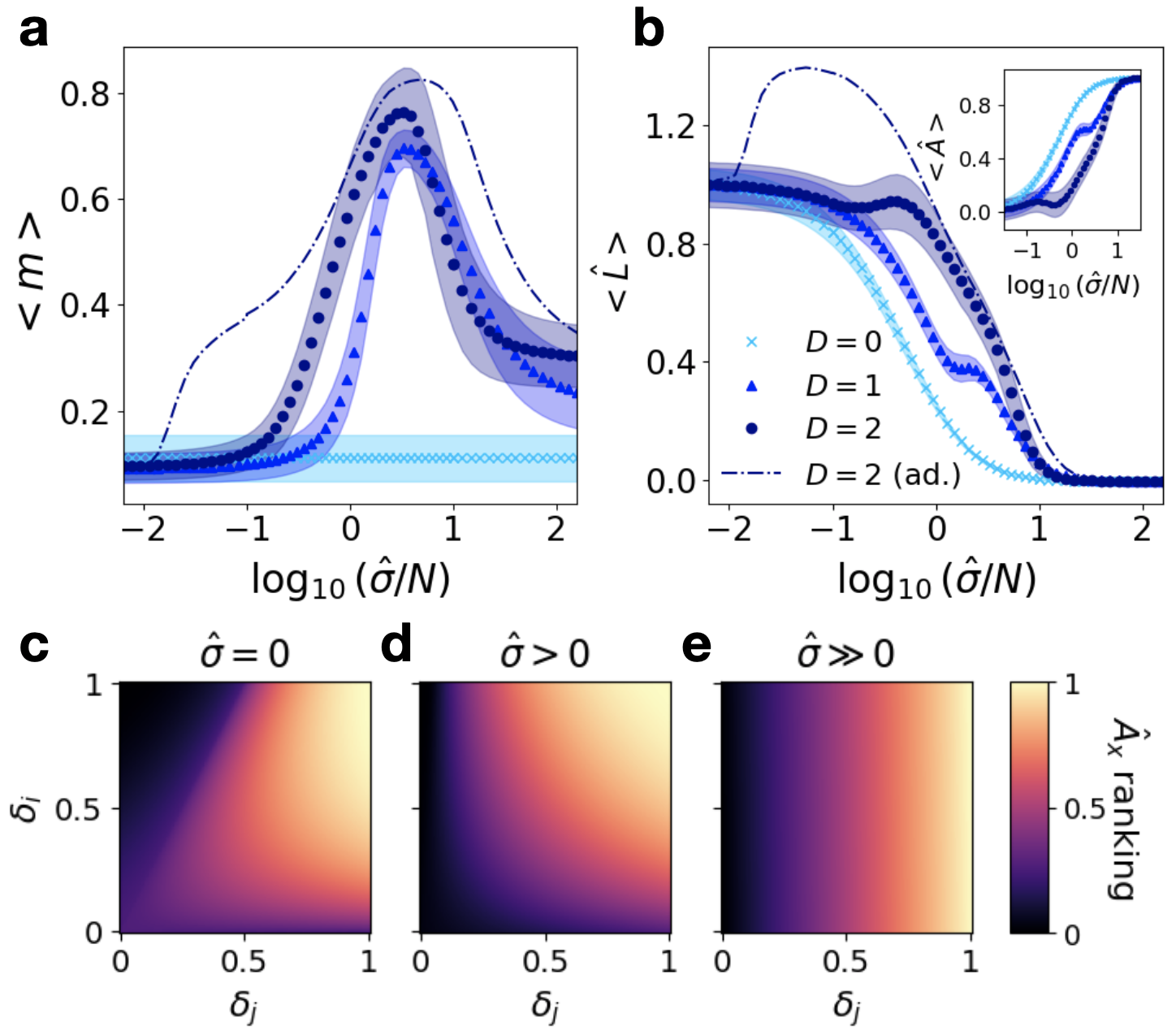}
  \caption{\textbf{Theoretical predictions.} \textbf{a:} Mean magnetization as a function of coupling, for increasing $D$ (from light to dark blue). The theory predicts disorder-order-disorder transitions in $\langle m \rangle$ for $D > 0$. \textbf{b:} Mean effective loss as a function of coupling (inset shows $\langle \hat{A} \rangle \approx 1 - \langle \hat{L}\rangle$ in a reduced range). The theory predicts a depth-induced delay which translates into a delayed critical point for the emergence of collective learning ($\langle \hat{A} \rangle > 0$). Markers denote mean values averaged over initial conditions and quenched disorders (shaded area represents one std.) and dashed lines the outcome of adiabatic protocols, averaged only over quenched sequences. \textbf{c:} Ranking of the $\hat{A}_x$ entries from Eq.(\ref{eq:crossaccuracy-main}) in the quenched `data' disorder plane for no coupling, \textbf{d:} Medium coupling and \textbf{e:} High coupling. See App. B for details on the numerical integration of the theory.}
  \label{fig2}
\end{figure}

\noindent {\bf Coarse-grained theory --}
We leverage on learning scale separation and results on deep linear networks \cite{saxe14,ziyin2022exact} --that shed light on the landscape of nonlinear networks and the existence of regularization-induced learning phase transitions-- to derive a coarse-grained theory describing the dynamics in Eq.(\ref{eq:NNdynamics}). Under suitable conditions (see Appendix section A.I for full derivation), namely linear activation functions, mean square loss, one-dimensional data and mean-field approximations, and considering a magnetization-like scalar order parameter $m_i$ (which coarse-grains the parameters of the $i$-th neural unit), the dynamics of Eq.(\ref{eq:NNdynamics}) project onto a reduced system of $N$ coupled differential equations 
\begin{equation}
    \dot{m}_i = \delta_i{m}_i^D -{m}_i^{2D+1} - \hat{\gamma} {m}_i + \frac{\hat{\sigma}}{N} \sum_{j = 1}^N q_{ij}({m}_{j}-{m}_{i}), 
    \label{eq:eff_model} 
\end{equation}
where $\dot{m}_i$ indicates the time derivative, $\hat{\sigma}$ and $\hat{\gamma}$ are effective hyper-parameters (coupling and regularization, respectively), $q_{ij}$ are the entries of the supra-network adjacency matrix and $D$ is the neural depth. The full information on the training set is encapsulated in $\bm{\delta}$, a sequence of scalars $\delta_i = \langle \hat{x}_i\hat{y}_i\rangle$, each of them being a one-dimensional projection of the expected input-output correlation, averaged over the data points assigned to the $i$-th unit. We build the effective test set by aggregating the private training distributions, such that $\langle \delta \rangle = (1/N)\sum_i \langle \hat{x}_i\hat{y}_i \rangle$. Interestingly,  the mean effective loss --evaluated on the test set and averaged over the units-- scales with the moments of the magnetization as 
\begin{equation}
\langle \hat{L} \rangle \sim \langle m^{2(D+1)} \rangle - 2 \langle \delta \rangle \langle m^{(D+1)}  \rangle, 
    \label{eq:eff_loss} 
\end{equation}
up to a constant that only depends on the dataset and can be neglected (see App. A.II for full derivation of Eq.(\ref{eq:eff_loss})). Our goal is to assess the onset of collective learning by solving Eq.(\ref{eq:eff_loss}) in the stationary regime of Eq.(\ref{eq:eff_model}). From now on, we take an all-to-all unweighted supra-network (i.e. mean-field) with $q_{ij} = 1$, $\forall \ i \neq j$. We shall distinguish three scenarios:
 
\medskip
\noindent \emph{Linear regression $({D = 0})$} -- When units have no hidden layers, Eq.(\ref{eq:eff_model}) becomes linear. Its stationary solution ($\dot{m}_i = 0$), found self-consistently, reads 
\begin{equation}
    m_i^* = \frac{\delta_i + \hat{\sigma}\langle \delta \rangle (1+\hat{\gamma})^{-1}}{1+\hat{\gamma}+\hat{\sigma}}.
    \label{eq:eff_d0} 
\end{equation}
The order parameter $\langle m \rangle = \langle \delta \rangle/(1+\hat{\gamma})$ is independent of coupling $\hat{\sigma}$, as shown in Fig. \ref{fig2}\textbf{a} (light blue crosses). Increasing $\hat{\sigma}$ only makes the distribution of magnetization narrower, which translates into a monotonously decreasing loss function, as displayed in Fig. \ref{fig2}\textbf{b}.

\medskip
\noindent \emph{Shallow networks $({D = 1})$} -- Eq.(\ref{eq:eff_model}) becomes 
\begin{equation}
    \dot{m}_i = \left(\delta_i -\hat{\gamma}\right){m}_i -{m}_i^{3} + \frac{\hat{\sigma}}{N} \sum_{j=1}^N \left(m_j - m_i\right). 
    \label{eq:eff_d1} 
\end{equation}
Remarkably, this equation is formally identical to the mean-field, zero-temperature version of the Ginzburg-Landau (GL) model with multiplicative quenched disorder ($\phi^4$ model \cite{van1994mean,buceta01,toral06,komin10,hohenberg14}, traditionally used to explore the critical behavior of condensed-matter systems with impurities under Landau's free energy approach. The local magnetization $m_i$ or \emph{spin} is the average parameter of our neural unit, whereas the heterogeneous allocation of training data plays the role of multiplicative quenched disorder, i.e. that of the material's random impurities \cite{van1994mean}. Eq.(\ref{eq:eff_d1}) displays a rich phenomenology which can be reinterpreted in the context of our neural system: For low coupling $\hat{\sigma}$, units trained on data $\delta_i > \hat{\gamma}$ relax in a double well local potential (with two symmetric stable points at $m_{i,\pm}^* = \pm \sqrt{\delta_i-\hat{\gamma}}$), and units with $\delta_i < \hat{\gamma}$ have a single equilibrium point at $m_i^* = 0$. This distribution produces a disordered state (with $|\langle m \rangle| \approx 0)$. At a critical coupling strength, a \emph{symmetry-breaking} mechanism induces a collective ordered phase (with $|\langle m \rangle| > 0)$ via a second-order collective phase-transition. For even larger coupling, the system returns to a `disordered' phase (via a reentrant phase-transition \cite{buceta01}), where all units become 
$m_i^* \approx 0 \ \forall \ i$, by means of a \emph{collective regularization} mechanism (when diffusion dominates in Eq.(\ref{eq:eff_model})). These disorder-order-disorder phase transitions are visualized in Fig. \ref{fig2}\textbf{a} (blue triangles). Interestingly, the critical behavior of the local magnetization triggers a non-monotonous decay of the effective loss. As shown in Fig. \ref{fig2}\textbf{b} (and inset), the non-monotonous shape delays the transition to the collective learning regime with respect to $D = 0$. 

\medskip 

\noindent \emph{Deep networks $({D > 1})$} -- In the deep learning realm, Eq.(\ref{eq:eff_model}) represents an exotic deformation of the GL model --which now can be seen as a $\phi^{2D+2}$ instead of a $\phi^4$ model-- with odd powers of the order parameter entering in Landau's free energy for even neural unit depths $D$, hence breaking rotational symmetry \cite{hohenberg14}. Fig. \ref{fig2}\textbf{a} (dark blue circles) shows that disorder-order-disorder transitions are still found for $D=2$, whereas the non-monotonic behavior of the effective loss is enhanced with respect to the shallow case. This effect translates into a further delay of the transition to the collective learning regime, as observed in Fig. \ref{fig2}\textbf{b} and its inset. Furthermore, the increased non-linearity of the deep case induces an effective landscape with many local minima. A simple linear stability analysis reveals that $m_i^* = 0$ is always a local minimum, which triggers a first-order, regularization-induced, phase transition at the single unit level (see \cite{ziyin2022exact} for details). Indeed, the local bistability gets further amplified at the collective scale. Figs. \ref{fig2}\textbf{a} and \ref{fig2}\textbf{b} show significant deviations between non-adiabatic (circles) and adiabatic (dashed line) protocols, which confirm that for $D > 1$ there is a strong sensitivity to initial conditions and a widespread presence of multistability. 

\noindent{\emph{Microscopic learning path} --} We finally leverage our theory to unveil the order in which neural units learn from each other as the system enters the collective learning phase. To tackle this problem analytically, we first construct the cross-loss matrix $\hat{L}_x$, with $(\hat{L}_x)_{ij} \sim m_i^{D+1}(m_i^{D+1}-2\delta_j)$. In the uncoupled regime of Eq.(\ref{eq:eff_model}), we have $m_i(D) \sim m_i(0)^{1/(D+1)}$ and $\hat{L}_x$ can therefore be estimated directly from the $D=0$ case. We extend this depth-independence as an \emph{ansatz} to the whole coupled regime and compute $\hat{L}_x$ using Eq.(\ref{eq:eff_d0}). Since a high accuracy requires a low loss \cite{goodfellow16}, 
we use $\hat{A}_x \sim -\hat{L}_x$ to get
\begin{equation}
(\hat{A}_x)_{ij} \sim \frac{\delta_i+\hat{\sigma}\langle \delta \rangle (1+\hat{\gamma})^{-1}}{1+\hat{\gamma}+\hat{\sigma}}\left(2\delta_j -\ \frac{\delta_i+\hat{\sigma}\langle \delta \rangle (1+\hat{\gamma})^{-1}}{1+\hat{\gamma}+\hat{\sigma}}\right). 
\label{eq:crossaccuracy-main}
\end{equation}
Eq.(\ref{eq:crossaccuracy-main}) presents a rich and interesting structure. Let us set $\gamma, \ \langle \delta \rangle \rightarrow 0^{+}$ and evaluate the normalized rankings of $\hat{A}_{ij}$ in the positive support of $\bm{\delta}$. First, in the uncoupled regime we have $(\hat{A}_x)_{ij} \sim 2\delta_i\delta_j -\delta_i^2$ (Fig. \ref{fig2}\textbf{c}). The approximate diagonal shape indicates that units have the highest cross-accuracy when evaluated on data $\delta_j$ that is similar to their training data $\delta_i$, as expected in the local learning regime. As coupling increases, $(\hat{A}_x)_{ij} \sim (2\delta_i \delta_j + \hat{\sigma}\langle \delta \rangle \delta_j)/(1+\hat{\sigma})$. The first symmetric term (Fig. \ref{fig2}\textbf{d}) dominates for a wide range of couplings. Instead, for very large coupling (Fig. \ref{fig2}\textbf{e}), the cross-accuracy depends only on the data evaluated, not on the unit predicting it, as expected when the units become so similar (due to diffusive coupling) that they make the same predictions and errors. Last, from Eq.(\ref{eq:crossaccuracy-main}) we estimate the cumulative cross-accuracy matrix $S_x$ to quantify the learning path as coupling increases. The integral $\hat{S}_x = \int_0^{\sigma'} \hat{A}_x d\sigma$ scales, in matrix form, as $\hat{S}_x \sim \log \sigma'\boldsymbol{\delta}\boldsymbol{\delta}^{\top}+\sigma'\langle \delta \rangle \boldsymbol{\delta}{\mathbf{1}}^{\top}$ (where $\boldsymbol{\delta}^{\top}$ is the transpose of the quenched `data' disorder vector and $\mathbf{1}$ is a vector of $N$ ones). Consistent with the low-dimensional nature of our theory, this calculation predicts that the learning path between the units is approximately rank-one, i.e. driven by a single effective dimension (the vector $\boldsymbol{\delta}$). From the previous analysis, the interpretation of $\boldsymbol{\delta}$ becomes clear: the higher the $\delta_j$ (the stronger the input-output correlation of the $j$-th data), the earlier the units will predict it well (which is amplified if the unit making the predictions is also trained on a high $\delta_i$). These findings are reminiscent of how learning works in a single neural network trained on global data \cite{saxe14}, which draws an interesting parallel between local and collective scales. 

\medskip

\noindent{\bf Validation --} We tackle the standard MNIST image classification task with $N=10$ coupled nonlinear feed-forward networks (each unit is trained on images of a unique class, a digit from 0 to 9). The system learns under the update rule in Eq.(\ref{eq:NNdynamics}), and performance is evaluated on an independent, multiclass test set using Eq.(\ref{eq:NNloss1}) and related metrics. We refer to the App. C for further details on the dataset (sec. C.I), neural architecture (C.II) and learning algorithm (C.III) used in the experiments. 
\begin{figure}[h]
  \includegraphics[width=\columnwidth]{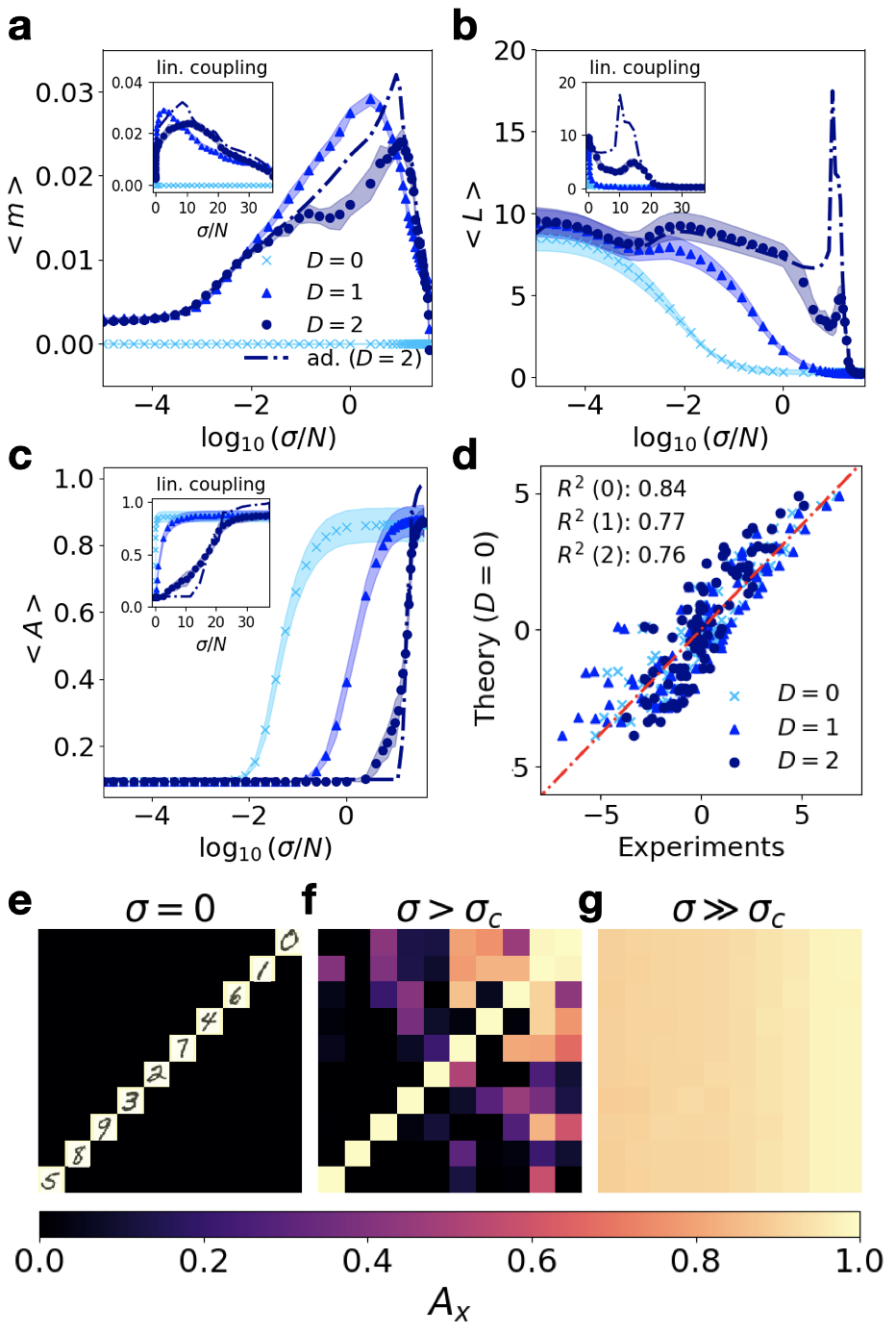}
  \caption{\textbf{MNIST experiments. a:} Mean magnetization, \textbf{b}: Mean test loss and \textbf{c}: Mean test accuracy as a function of coupling (in log scale), for depths $D=0,1,2$. Shaded area shows one std. over 10 independent runs and inset shows results in a linear scale of coupling. The dashed line is the outcome of adiabatic protocols $(D = 2)$. \textbf{d}: Correlation between the entries of the empirical $S_x$ matrix (averaged over 10 runs and shifted by the mean) for $D = 0,1,2$ against the entries of the rank-one truncation of $S_x$ for $D = 0$. We show the $R^2$ for the three cases and the best linear fit as a red line. \textbf{e}: Empirical $A_x$ for $D = 1$ at $\sigma = 0$ , \textbf{f}: $\sigma = 3$ and \textbf{g}: $\sigma = 30$. Rows and columns have been ordered from lowest to highest cross-accuracy (as indicated by MNIST digits in \textbf{e}) and the y-axis is inverted to aid the visual comparison with Fig. \ref{fig2}.}
  \label{fig3}
\end{figure}
\noindent Fig. \ref{fig3} shows results for varying neural depth $D$, across a wide range of coupling values and averaging over independent realizations of the process. Fig. \ref{fig3}\textbf{a} confirms that, for $D > 0$, disorder-order-disorder phase transitions in the collective order parameter $\langle m \rangle$ appear as coupling increases. As predicted, Fig. \ref{fig3}\textbf{b} shows that the phase transitions induce a non-monotonous behavior of the mean loss, along with a delay of the loss decay to zero (which is amplified with neural depth $D$). This effect triggers the appearance of a (depth-delayed) critical point where collective learning emerges, as observed in Fig. \ref{fig3}\textbf{c}. The mean accuracy metric grows abruptly from $\langle A \rangle \approx 0.1$ (local learning phase) where units only predict well the class belonging to their private training set, to $\langle A \rangle > 0.1$ at the critical point, up to $\langle A \rangle \approx 1$ (collective learning phase) for larger coupling, where units predict well all classes. Also note that for $D = 2$ the differences between non-adiabatic (circles) and adiabatic protocols (dashed line) confirm the presence of multistability in the deep case: Intriguingly, the collective learning transition is abruptly delayed when the system has `memory' (i.e. when the parameters are only initialized at $\sigma = 0$). Fig. \ref{fig3}\textbf{d} displays the empirical correlation between the entries of the cumulative cross-accuracy matrix and its rank-one approximation for $D = 0$ (the approximate learning path predicted by the theory) which is computed using the singular value decomposition of $S_x$ and keeping only the leading term. A considerable high $R^2$ score is sustained for varying depth, providing a quantitative validation of the coarse-grained approach at the microscopic level. Finally, Fig. \ref{fig3} (bottom) shows snapshots of the cross-accuracy matrix $A_x$ (for $D = 1$) at three values of coupling strength. In \textbf{e}: a local phase with an associated diagonal $A_x$ for low coupling (as Fig. \ref{fig2}\textbf{c}), \textbf{f}: a symmetric collective learning phase for medium coupling --above the critical point-- (as Fig. \ref{fig2}\textbf{d}) and \textbf{g}: a column-dependent phase for large coupling \cite{footnote1}, where accuracy only depends on the class that is evaluated (as Fig. \ref{fig2}\textbf{e}), thus reproducing the three patterns predicted by our analysis. 

\medskip

\noindent {\bf Discussion --} 
While our effective --coarse-grained-- theory is based on a list of simplifications and approximations, the predictions about the onset and properties of collective learning are well confirmed on realistic (high-dimensional, highly non-linear) experiments on the MNIST dataset, which supports the mechanistic interpretability of the collective learning paradigm. Yet, much work is required to better understand its rich phenomenology and the relation of decentralized learning schemes with the statistical physics of (deformed) Ginzburg-Landau models. Note that our proof of concept did not cover the myriad of learning phases found in isolated neural networks \cite{liu22grokking,jacot18,martin21,achille19,sclocchi23}, which can get more exotic in the collective case. Extending the effective theory to the complex plane may capture some of these nuances and even increase the analytical tractability (thanks to exact dimensional reductions and other techniques available for coupled oscillators \cite{ott08,bick20,gottwald15,arola21}). Other promising directions include exploring how the shape and difficulty of a dataset \cite{sclocchi23} (\emph{effectively} encapsulated in the distribution of quenched disorders) and the supra-network topology \cite{newman10} (beyond the mean-field, undirected, pair-wise and time-independent case) affect the nature of the phase transitions \cite{gardenes11,skardal15a,arola22}, and refining the diffusive coupling mechanism to model heterogeneous neural learners and indirect communication channels \cite{li19,chatterjee12,gao22}. 

In a nutshell, this work offers a mathematical foundation for collective learning in natural and artificial systems. Our perspective enriches deep learning theories \cite{carleo2019machine,mezard23,liu22grokking,ziyin2022exact,mehta19,roberts22} and statistical physics approaches to interacting brains \cite{sole19,pinero19}, and makes a first step towards a next-generation type of physical models \cite{bick20,gomezgardenes18} to describe emergent social behavior --such as collective learning-- in populations of interacting agents. Last, it has not escaped our notice that this framework could contribute to tackle the so-called alignment problem \cite{vinuesa2020role} when independent learning models are put in interaction.
 
\medskip

\noindent {\bf Acknowledgments:} We thank many colleagues at IFISC, C. Granell, A. Arenas and A. Raya for useful discussions. We acknowledge funding from project DYNDEEP (EUR2021-122007) from the Agencia Estatal de Investigaci\' on MCIN/AEI/10.13039/501100011033. LL additionally acknowledges funding from project MISLAND (PID2020-114324GB-C22), and María de Maeztu project CEX2021-001164-M.\\


%

\appendix

\section{Analytical results}
\subsection{Derivation of the effective dynamics}
The dynamics of the collective learning algorithm introduced in Eq.(2) reads as  
\begin{equation}
    \theta^{\alpha}_{i}(t+1) = \theta^{\alpha}_{i}(t)  -\eta\nabla_{\theta^{\alpha}_{i}}  {\cal{{L}}}_i
     + \frac{\eta \sigma}{N} \sum_{j = 1}^N q_{ij} \left(\theta^{\alpha}_{j}(t)-\theta^{\alpha}_{i}(t)\right), 
    \label{eq:NNdynamics_supp}
\end{equation}
where $\theta$ is a parameter of a neural unit ($\alpha$ runs over the number of parameters and $i$ over the number of units in the ensemble, $N$) and $t$ denotes the iteration or step. $\eta$ is the learning rate, $\sigma$ the coupling strength, $q_{ij}$ the entries of the adjacency matrix of the supra-network, and $\nabla$ is the gradient operator acting on the local loss. 

To derive our coarse-grained version of this model, we begin by assuming that the learning rate $\eta$ in Eq.(\ref{eq:NNdynamics_supp}) is sufficiently small, such that we can approximate the discrete dynamics by the continuous version
\begin{equation}
    \dot{\theta}_i^\alpha = - \nabla_{\theta^{\alpha}_{i}}  {\cal{{L}}}_i(\bm{x}_i,\bm{y}_i,\bm{\theta}_i,\gamma) +  \frac{\sigma}{N} \sum_{j = 1}^N q_{ij} (\theta^{\alpha}_{j}-\theta^{\alpha}_{i}),
    \label{eq:eff_dynamics_supp} 
\end{equation}
where $\dot{\theta}^\alpha_i = d \theta^\alpha_i/ dt$ is the time-derivative of a parameter, the time step $dt$ in the numerical integration of Eq.(\ref{eq:eff_dynamics_supp}) corresponds to the learning rate $\eta$ and bold variables denote vectors (a notation used also in the main text). 

Now we follow a common procedure in physics, i.e. to linearize the system and study the properties of the resulting linear approximation \cite{mehta19,saxe14,ziyin2022exact}. When the choice of $\ell(\bar{y},y)$ is the mean square error (MSE), the loss function of a deep linear network can be written as \cite{ziyin2022exact}
\begin{gather}
    {\cal{\hat{L}}}_i = \mathbb{E}_{x_i} \left[\left( \sum_{d_o,\dots,d_{D+1}}^{n_o,\dots,n_{D+1}} [\Theta_{d_{D+1}d_{D}}^{D+1}]_i\dots [\Theta_{d_1d_0}^{1}]_i[x_{d_0}]_i - y_i \right)^2 \right] \nonumber \\
    + \gamma \sum_{d = 1}^{D+1} || \Theta_i^{d}||_2,
    \label{eq:eff_loss1} 
\end{gather}
where $\mathbb{E}_{x_i}[\cdot]$ denotes the expected value over the accessible data points and $|| \Theta_i^{d}||_2$ is the squared $L_2$ norm of all elements in the parameters' matrix of the $d$-layer. Each sum in the first term of Eq.(\ref{eq:eff_loss1}) runs over all the neurons of a given layer with width $n_d$. The index $i$ is kept to reflect that the loss, the parameters and the data belong to the $i$-th neural unit in the ensemble setting. 

Here we follow the `mean-field' analysis of \cite{ziyin2022exact}, assuming that both the input and output are one-dimensional, and we approximate each matrix $\Theta_i^d$ by the mean value of its entries, a scalar $c_d\hat{\theta}_i^d$, where $c_d$ is a layer-dependent constant. For the sake of simplicity, we set $c_d = 1$ in this study. The loss function of the $i$-th unit in Eq.(\ref{eq:eff_loss1}) then reads as 
\begin{gather}
    {\cal{\hat{L}}}_i \approx \mathbb{E}_{x_i} \left[\left(\hat{x}_i \prod_{d=1}^{D+1}\hat{\theta}_i^d -\hat{y}_i \right)^2 \right] + \gamma\sum_{d = 1}^{D+1}  (\hat{\theta}_i^d)^2,
    \label{eq:eff_loss2} 
\end{gather}
where $\hat{x}_i$ and $\hat{y}_i$ are the one-dimensional projections of input and output for a data point assigned to the $i$-th unit. Under this mean-field approximation, we can reduce the number of coupled equations in Eq.(\ref{eq:eff_dynamics}) from a very large number ($\alpha$ runs from 1 to the dimension of $\mbox{vec}(\cup_{d=1}^{D+1}\Theta^d)$) to just $(D+1)\times N$ equations, one for each layer of parameters in each unit. In particular, we can now compute the gradient in Eq.(\ref{eq:eff_dynamics}) with respect to $\hat{\theta}_i^d$, and using the loss function of Eq.(\ref{eq:eff_loss2}) we get 
\begin{equation}
    \dot{\hat{\theta}}_i^d = - \frac{\partial {\cal{\hat{L}}}_i}{\partial \hat{\theta}_i^d} +  \frac{\sigma}{N} \sum_{j = 1}^N q_{ij} (\hat{\theta}^{d}_{j}-\hat{\theta}^{d}_{i}),
    \label{eq:eff_dynamics2} 
\end{equation}
with $i \in [1,N]$, $d \in [1,D+1]$ and 
\begin{equation}
    \frac{\partial {\cal{\hat{L}}}_i}{\partial \hat{\theta}_i^d} = -2\langle \hat{x}_i\hat{y}_i\rangle \prod_{k \neq d}^{D+1}\hat{\theta}_i^k + 
    2 \langle \hat{x}_i^2\rangle \hat{\theta}_i^d\prod_{k \neq d}^{D+1}(\hat{\theta}_i^k)^2  + 2\gamma \hat{\theta}_i^d,
    \label{eq:eff_gradient} 
\end{equation}
where we use the simplified notation $\langle \cdot \rangle$  to indicate an average (here over the data points assigned to a unit) as we do in the main text. Let us define the local magnetization $m_i = (D+1)^{-1}\sum_d \hat{\theta}_i^d$ as the average learning parameter of the $i$-th unit and compute the average value of the gradient in Eq.(\ref{eq:eff_gradient}), summing over the different layers, leading to 
\begin{equation}
   \frac{\partial {\cal{\hat{L}}}_i}{\partial m_i} \approx  -2\langle \hat{x}_i\hat{y}_i\rangle m_i^D + 2\langle \hat{x}_i^2\rangle m_i^{2D+1} + 2 \gamma m_i, 
    \label{eq:eff_gradient2} 
\end{equation}
where we have used the mean-field approximations
\begin{gather}
(D+1)^{-1}\sum_{d=1}^{D+1} (\partial {\cal{\hat{L}}}_i /\partial \hat{\theta}_i^d) \approx \frac{\partial{{\cal{\hat{L}}}_i({m_i})}}{\partial{m_i}},\\
(D+1)^{-1}\sum_{d=1}^{D+1} \prod_{k \neq d}^{D+1} \hat{\theta}_i^k \approx m_i^D,\\
(D+1)^{-1}\sum_{d=1}^{D+1} \hat{\theta_i}^d \prod_{k \neq d}^{D+1} (\hat{\theta}_i^k)^2 \approx m_i^{2D+1}.
\end{gather}
Now we plug Eq.(\ref{eq:eff_gradient2}) into the sum of the $(D+1)$ equations in Eq.(\ref{eq:eff_dynamics2}) for each unit. Assuming orthogonal input representations \cite{saxe14,jacot18} (which holds exactly for whitened input data \cite{saxe14}) we set $\langle \hat{x}_i^2 \rangle = 1$ without loss of generality. After re-scaling time with $dt' = 2dt$ and absorbing any additional constants into the effective variables $\hat{\sigma}$, $\hat{\gamma}$ and $\delta_i = \langle \hat{x}_i\hat{y}_i\rangle$, we get 
\begin{equation}
    \dot{m}_i = \delta_i{m}_i^D -{m}_i^{2D+1} - \hat{\gamma} {m}_i + \frac{\hat{\sigma}}{N} \sum_{j = 1}^N q_{ij}({m}_{j}-{m}_{i}), 
    \label{eq:eff_model_supp} 
\end{equation}
which corresponds to Eq.(4) in the main text.

\subsection{Derivation of the effective loss}

To derive the mean effective loss, first note from Eq.(\ref{eq:eff_loss2}) that the effective output function of a neural unit takes the simple form $\bar{y}_i \approx m_i^{D+1}\hat{x}_i$. Then, using the MSE as the individual loss function $\ell(\bar{{y}}_i,{y}_i)$ --where $\bar{y}_i$ and $y_i$ are both scalars--, the effective loss of the $i$-th unit when evaluating data assigned to the $j$-th unit (i.e. the cross-loss entry $(L_{x})_{ij}$) becomes 
\begin{equation}
(\hat{L}_x)_{ij} = \left(m_i^{D+1}\hat{x}_j-\hat{y}_j\right)^2. 
\end{equation}
Decomposing terms, neglecting the constant $\langle \hat{y}_j^2\rangle$ that only depends on the dataset, and assuming again orthogonal input data, we have
\begin{equation}
(\hat{L}_x)_{ij} \sim m_i^{D+1}(m_i^{D+1} - 2\delta_j), 
\end{equation}
which is used as the starting point to estimate the microscopic learning path. Finally, by summing all the entries of this matrix and using that in the test set $\langle \delta \rangle = N^{-1}\sum_j \delta_j$, we get 
\begin{equation}
\langle \hat{L} \rangle \sim \langle m^{2(D+1)} \rangle - 2 \langle \delta \rangle \langle m^{(D+1)}  \rangle, 
    \label{eq:eff_loss_supp} 
\end{equation}
which corresponds to Eq.(5) in the main text.

\section{Numerical integration}

We integrate the dynamics of Eq.(\ref{eq:eff_model_supp}) using a RK45 method implemented with the SciPy Python solver. At each value of the coupling strength, we iterate for a time span of $10 s$, averaging the metrics during the second half of the simulation to discard transient behavior. The number of nodes is set to $N = 200$, regularization is $\hat{\gamma} = 10^{-3}$ and results are averaged over 200 realizations. The sequence of quenched disorders $\bm{\delta}$ is drawn from a Gaussian distribution $\mathcal{N}(0,2)$ and the initial condition for the magnetization values is the uniform support in $[-2+\epsilon,2+\epsilon]$, where a small $\epsilon$ is used to break the symmetry towards positive solutions of $\langle m \rangle$ and reproduce what occurs in the experiments (where the nonlinear neuron activation function always produce $\langle m \rangle > 0$ in the ordered regime). 

We consider two protocols in order to show the presence of multistability and the dependence on the initial conditions. In the adiabatic protocol (dashed lines in the main text) we use the distribution of magnetization values in the units from the previous coupling value as the initial condition of the following one, while in the non-adiabatic protocol (markers) we initialize the magnetization values at each coupling strength. 

\section{Experimental details} 
Here we describe the dataset, the architecture of the neural units and the parametrization of the learning algorithm used in our experiments. 
\subsection{Dataset} 
We use the standard MNIST dataset to validate the effective theory. It consists of 60000 labelled black-and-white images of $28 \times 28$ pixels corresponding to handwritten digits from zero to nine, and it is divided into train set (50000) and test set (10000) for cross-validation purposes. The images are flattened as vectors to feed as inputs into the neural network (see below). 
\subsection{Neural architecture}
We use fully-connected feed-forward neural networks as the backbone of the neural units. We remark that the current version of the algorithm can be implemented on any other neural architectural (e.g. convolutional nets, recurrent nets or transformers) as long as the number of parameters is the same in all the units. For each data point $x \in \mathbb{R}^{n_0}$, we use $h^d(x) \in \mathbb{R}^{n_d}$ and $x^d(x) \in \mathbb{R}^{n_d}$ for the pre- and post-activation functions. The recurrence relation for a layer of the network is defined as 
\begin{equation}
    h^{d+1} = x^{d}W^{d+1}+b^{d+1}, \mbox{ } x^{d+1} = \phi(h^{(d+1)}),
    \label{eq:NNactivation}
\end{equation}
where $\phi()$ is a point-wise, non-linear activation function (we use the standard `ReLU' function $\phi(x) = max(0,x)$ \cite{goodfellow16}), $W^{d+1} \in \mathbb{R}^{n_d \times n_{d+1}}$ is the matrix of weights between layers and $b^{d+1} \in \mathbb{R}^{n_d+1}$ is the vector of biases. We define $\Theta^d$ as the matrix of layer $d$ that includes both the weights and biases, thus $\Theta^d \in \mathbb{R}^{(n_d+1) \times (n_{d+1}+1)}$. We also define $\theta = \mbox{vec}(\cup_{d=1}^{D+1}\Theta^d)$, as the flattened vector of all the parameters of a single neural unit. We fix the number of hidden neurons to $n_d = 20$ for all hidden layers.

\subsection{Training and evaluation} To minimze the local term in Eq.(\ref{eq:NNdynamics}), we use standard Backpropagation with vanilla mini-batch gradient descent (without momentum), with batch size $B = 2^5$ and Cross-Entropy as the loss function \cite{goodfellow16}, suitable for categorical outputs. The batch size is sufficiently large \cite{sclocchi23} to neglect for simplicity an additive noise term that would emerge in Eq. (\ref{eq:eff_model_supp}) if SGD was used to approximate the global loss function by taking the gradient of a single data point. The learning rate is fixed to $\eta = 0.005$ and weight decay to $\gamma = 10^{-3}$. Results for each coupling value $\sigma$ are time-averaged during $2\cdot 10^4$ batch iterations, after $2 \cdot 10^4$ more iterations to discard transient behavior. We run independent realizations where parameters are initialized with $\mathcal{N}(0,1)$.\\

\noindent {\bf Data and code availability:} Python scripts (based on PyTorch library) will be available on github.com/mystic-blue/collective-learning upon publication.

\end{document}